\documentclass[twocolumn,showpacs,preprintnumbers,aps]{revtex4}

\usepackage{dcolumn}
\usepackage{bm}
\usepackage{amssymb}

\newif\ifpdf
\ifx\pdfoutput\undefined
\pdffalse 
\else
\pdfoutput=1 
\pdftrue
\fi

\ifpdf
\usepackage[pdftex]{graphicx}
  \usepackage[pdftitle=
		   {The black hole tunnel phenomenon}
		   pdfauthor={Andreas de Vries},%
		   pdfpagemode=FirstPage,%
            colorlinks=false,hyperindex=true,plainpages=false,%
           ]{hyperref}
  \DeclareGraphicsExtensions{.pdf}
\else
  \usepackage{graphicx}
  \DeclareGraphicsExtensions{.eps,.epsf}
\fi

\newtheorem{satz}{Theorem}
\newtheorem{lem}[satz]{Lemma}

\begin{document}

\title
{The black hole tunnel phenomenon}
\author{Andreas de Vries}
 \altaffiliation{AG Mathem.\ Physik, Ruhr-Universit\"at, D-44780 Bochum}
\affiliation{FH S\"udwestfalen, 
University of Applied Sciences, 
Haldener Stra{\ss}e 182,
D-58095 Hagen, Germany}
\email{de-vries@mfh-iserlohn.de}
\author
{Theodor Schmidt-Kaler}
\affiliation{Georg-B\"uchner-Stra{\ss}e 37, 
D-97276 Margetsh\"ochheim, 
Germany}
 \altaffiliation{Astronom. Institut, Ruhr-Universit\"at, D-44780 Bochum}
\date{\today}

\begin{abstract}
The potentials of spin-weighted wave 
equations in various Kerr-Newman black holes are analyzed.
They all form singular potential barriers at the event horizon.
Applying the WKB approximation it is shown that no particle 
can tunnel out of the interior of
a static black hole. 
However, 
photons inside a non-extremely rotating Kerr black hole
may tunnel out into the outer space, whereas neutrinos, electrons,
and gravitons may not. 
If the rotation is extremal,
any particle may tunnel out, under restrictive conditions.
It is unknown 
whether photons and gravitons may tunnel out if the
black hole is charged and rotating.
\end{abstract}


\maketitle

\section{Introduction\label{sec-intro}}
In classical general relativity a causal particle inside the event 
horizon is inevitably pulled towards the center of the black hole, 
at least until it reaches the Cauchy horizon representing a barrier to 
predictability \cite{Israel-1992}. In particular, no classically 
relativistic particle, be it of positive or vanishing rest-mass, can escape 
from the black hole interior. However, Hawking \cite{Hawking-1976}
has shown that photons in effect can leave the black hole if a 
quantized photon field in the curved spacetime is supposed, 
cf.\ \cite{Wald-1984}.

In general, quantization of
spin-weighted waves in a curved spacetime
yields the notion of spin particles in a gravitational field. 
As such, spin-weighted waves are the basis
for semiclassical quantum gravity. Their properties and
behavior in the outer space of Kerr black holes
have been extensively studied
\cite{Teukolsky-1973,Starobinsky-Churilov-1974,
Unruh-1974,Gueven-1977,Chandrasekhar-1983}, 
and could in part even be
extended to charged Kerr-Newman black holes
\cite{Page-1976,Lee-1977,de-Vries-1996,de-Vries-1995}.
By the symmetries of these spacetimes the equations turn out to be
separable in special coordinate frames, thus being
mathematically tractable to a certain extent. 

In the framework of classical general relativity the examination of waves
in the outer space of a black hole seemed sufficient, since the
event horizon acts like a perfect semi-permeable
membrane, letting
in any form of energy and matter but allowing none to get out.

In the present paper we extend these considerations to the region beyond
the event horizon. Topologically, this means nothing particular, since
the spacetime curvature remains finite in this domain and the
event horizon is nothing more than a coordinate singularity.
As expected, the event horizon causes a singular repulsive 
potential in the respective wave equations. However, more 
detailed examination
shows the remarkable, and to our knowledge yet 
unmentioned, property of the potential barrier to be singular enough
to prevent nearly all kinds of spin-weighted waves from tunneling through
it --- unless electromagnetic waves. 

Our analysis below shows that in fact a photon inside an uncharged rotating 
black hole may tunnel through the event horizon to the outside region,
but a graviton, a neutrino or an electron may not.
If the black hole is non-rotating, no particle at all
can tunnel out of it. In case of a rotating and electrically charged black hole
it remains still unknown what kind of particle may tunnel out,
it is only sure that electrons and neutrinos \emph{may not}.

The present paper is organized as follows. In section \ref{sec-KN} we introduce
the basic notation and properties of Kerr-Newman spacetimes. In section
\ref{sec-spin-waves} we analyze the equations for massless waves
in Kerr geometry and present a proof that only photons of certain discrete 
frequencies may tunnel out of a rotating but non-extremal Kerr black hole.
The case of extremal rotation is considered in section \ref{sec-rotation},
the mathematically not completely tractable case of massless waves in a 
charged Reissner-Nordstr{\o}m spacetime in section \ref{sec-RN}, and
the Dirac equation in a general Kerr-Newman
spacetime in section \ref{sec-Dirac}.
Finally, in section \ref{sec-discussion} we 
sum up the results and discuss them.

\section{Kerr-Newman spacetimes\label{sec-KN}}
Suppose a Kerr-Newman black hole with the three real parameters $M$, $a$, 
and $Q$. They are related to the mass $\mathfrak M$ (in kg), 
the angular momentum $\mathfrak J$ (in kg m$^2$ s$^{-1}$) and the 
electrical charge 
$\mathfrak q$ (in kg$^{1/2}$ m$^{3/2}$ s$^{-1}$) by the relations
\begin{equation}
	{\mathfrak{M}} = { c^2 M \over G }, \qquad
	{\mathfrak{J}} = ac{\mathfrak{M}}, \qquad 
	{\mathfrak{q}} = { c\, Q \over \sqrt{G} }.
	\label{parameters}
\end{equation}
Here $G$ is the gravitational constant, and $c$ the speed of light.
The non-vanishing contravariant components $g^{ij}$ of the metric tensor 
in Boyer-Lindquist coordinates $(x^0$, \ldots, $x^3)=(ct$, $r$, $\theta$,
$\varphi)$ are then
$$
 \begin{array} {c}  \displaystyle
    g^{tt} = { (r^2+a^2)^2 \over c^2 \rho \bar\rho \Delta} 
              - {a^2 \sin^2\theta \over c^2 \rho \bar\rho}\, ,\
    g^{rr} = - { \Delta \over \rho \bar\rho },\
    g^{\theta\theta} = - { 1 \over \rho \bar\rho },
                \\[2ex]  \displaystyle
    g^{t\varphi} = { (2Mr - Q^2) a \over \rho \bar\rho 
             \Delta},\quad
    g^{\varphi\varphi} 		= - {\Delta - a^2 \sin^2\theta \over \rho \bar\rho 
             \Delta
                  \sin^2\theta },
 \end{array} 
$$
($g^{t\varphi}=g^{\varphi t}$), where 
\begin{equation}
  \rho = r + \mbox i a \cos\theta, \qquad
  \Delta = (r-r_{+}) (r-r_{-}),
  \label{rho-Delta}
\end{equation}
with the \emph{event horizon} $r_{+}$ and the \emph{Cauchy horizon} $r_{-}$
given by
\begin{equation}
    r_{\pm} = M \pm \sqrt{M^2-a^2-Q^2}.
  \label{horizonte}
\end{equation}
Only the points of the set $\rho = 0$ have infinite Riemann curvature 
and thus locate the curvature singularity of the spacetime.
For $a\not= 0$, it in fact forms a ring of radius $|a|$ in 
the equatorial plane $\theta=\pi/2$, cf.~\cite{Chandrasekhar-1983}.
Because the \emph{cosmic censorship hypothesis} forbids naked singularities, 
the event horizon $r_{+}$ must necessarily exist. Thus the square root of
(\ref{horizonte}) must be real, i.e.
\begin{equation}
	a^2 + Q^2 \leqq M^2.
	\label{ccc}
\end{equation}

\section{Spin-weighted massless waves in Kerr geometry\label{sec-spin-waves}}
Let be $Q=0$, and define
$s$ $\in$ \{0, $\pm \frac12$, $\pm 1$, $\pm 2$\} as the \emph{spin-weight}. 
Then the sourcefree perturbation equations
for scalar ($s$ $=$ 0), two-component neutrino ($s$ $=$ $\pm \frac12$), 
electromagnetic
($s$ $=$ $\pm 1$), and gravitational fields ($s$ $=$ $\pm 2$) are given by 
wave equations which by the the symmetries of the Kerr spacetime, 
viz.\ stationarity and axialsymmetry,
admit the separable solutions
\begin{equation}
	\psi( t,r,\theta,\varphi) = R_s (r)\, S_s(\theta)\,
	  {\mathrm{e}}^{-{\mathrm{i}} \omega c t} \,
	  {\mathrm{e}}^{{\mathrm{i}} m \varphi} 
\label{separable solutions}
\end{equation}
with the constants $\omega\in(0,\infty)$ and $m$ $\in$ $\mathbb{Z}$. Here 
$R_s$ obeys the radial equation
\begin{eqnarray}
	\Delta^{-s} {{\mathrm{d}} \over {\mathrm{d}} r } 
	  \left( \Delta^{s+1} {{\mathrm{d}} R_s \over {\mathrm{d}} r } \right)
	  + \bigg( {K^2 - 2{\mathrm{i}} s (r-M) K \over \Delta } 
		\nonumber & \\* 
		  + 4 {\mathrm{i}} s\omega r - \lambda \bigg) R_s = 0,
\label{radial equation}
\end{eqnarray}
and $S_s$ solves the angular equation 
\begin{eqnarray}
	  {1 \over \sin\theta } {{\mathrm{d}} \over {\mathrm{d}} \theta } 
	  \left( \sin \theta\ {{\mathrm{d}} S_s \over {\mathrm{d}} \theta } \right)
	  + \bigg( (a\omega \cos \theta - s)^2
		\nonumber & \\* 
		- { (m + s \cos\theta)^2 \over \sin^2\theta }
		- s(s-1) + A \bigg) S_s = 0,
\label{angular equation}
\end{eqnarray}
with
\begin{equation}
	K = K(r) = (r^2 + a^2)\, \omega - am, 
\label{K-def}
\end{equation}
and $\lambda = A + a^2 \omega^2 - 2am\omega$,
cf. \cite{Teukolsky-1973,Starobinsky-Churilov-1974}
The constants
$\lambda$, $A$ $\in$ ${\mathbb{R}}$ are separation constants obtaining some discrete values
depending on the boundary conditions of $S_s$.
(For details see \cite{Futterman-et-al-1988}, \S2.1.)
To be more explicit, $\lambda$ $=$ 
${}^{}_s\lambda^m_l$ $=$ ${}^{}_sE^m_l$ $+$ $a^2\omega^2$ 
$-$ $s(s+1)$, where ${}^{}_sE^m_l$ $=$
$l(l+1)$ $-$ $2 a m \omega s^2 /(l(l+1))$ $+$ $O( (a\omega)^2 )$, with
$l$ $\geqq$ $\max(|s|, |m|)$ and
$l$ $\in$ $\mathbb{N}$ for integer spin $s$, $l$ half-odd integer
for $|s|$ $=$ 1/2,
cf.\ \cite{Press-Teukolsky-1973} eq.~(3.10).

\begin{satz}
\label{satz-potential}
	Let be given the operator $A_{(s)}:$ $C^2_0({\mathbb{R}},{\mathbb{C}})$ $\to$ 	$C^2_0({\mathbb{R}},{\mathbb{C}})$,
	\begin{equation}
		A_{(s)} = {{\mathrm{d}}^2 \over {\mathrm{d}}r^2} + q_s
	\label{A-s-def}
	\end{equation}
	with
	\begin{eqnarray}
		q_s =
		 { K^2 + (1 - s^2) (r-M)^2 \over \Delta^2 }
		 + { |s| - 1 - \lambda \over \Delta }
		\nonumber \\*
		 + {2 {\mathrm{i}} s \over \Delta }
		  \left( 2 \omega r - {(r-M)K \over \Delta} \right).
	\label{q-s-def}
	\end{eqnarray}
	With the functional transformation
	\begin{equation}
		u_s = \Delta^{(s + 1)/2} R_s
	\label{u-s-def}
	\end{equation}
	thes radial equation (\ref{radial equation}) is equivalent to 
	$A_{(s)}u_s$ $=$ $0$.
\end{satz}
\emph{Proof.}
First we see that 
with the functions $P_s$ given by the transformation
\begin{equation}
	P_s = \left\{ \begin{array}{lr}
	  \Delta^{s} R_{s} & \mbox{if } s \geqq 0, \\
	  R_{s} & \mbox{if } s < 0,
	  \end{array} \right.
\label{radial transformation}
\end{equation}
(i.e., $P_{|s|}$ $=$ $\Delta^{|s|} R_{|s|}$ and $P_{-|s|}$ $=$ $R_{-|s|}$)
the radial differential equation (\ref{radial equation}) is equivalent to 
$T_{(s)} P_s = 0$, with the differential operator
\begin{equation}
	T_{(s)} = {{\mathrm{d}}^2 \over {\mathrm{d}}r^2 } + g_s {{\mathrm{d}} \over {\mathrm{d}}r } + f_s,
\label{T-def}
\end{equation}
where
\begin{equation}
	f_s = { K^2 - 2{\mathrm{i}} s(r-M) K \over \Delta^2 }
	  + { 4 {\mathrm{i}} s\omega r - \lambda \over \Delta},
\label{f-s}
\end{equation}
and
\begin{equation}
	g_s = {2 (1-|s|) (r - M) \over \Delta}.
\label{g-s}
\end{equation}
(cf.\ \cite{Chandrasekhar-1983}, eqs.\ (3), (96), (97) in chapter 8). 
With
\begin{equation}
	{1\over 2} {{\mathrm{d}} g_s \over {\mathrm{d}}r} 
	 = { 1-|s| \over \Delta } \left( 1 - {2(r-M)^2 \over \Delta}\right),
\label{g'-s}
\end{equation}
we see that
$	q_s = f_s - {g_s^2 \over 4}
	 - \frac12 {{\mathrm{d}} g_s \over {\mathrm{d}}r}.
$
Since d$\Delta$/d$r$ $=$ $2(r-M)$, we have
$ \int g_s \, {\mathrm{d}}r 
    = (1-|s|) \ln \Delta.$
By elementary means (e.g.\ \cite{Hartman-1964}, p.~323) we deduce from this 
result that, by the transformation 
$
	u_s = \Delta^{(1 - |s|)/2} P_s,
$
the equation
$T_{(s)} P_s = 0$ is equivalent to $A_{(s)} u_s = 0$.
\hfill{$\square$}

\begin{figure} 
	\centering
	\begin{picture}(0,0)%
	 \includegraphics[width=50mm]{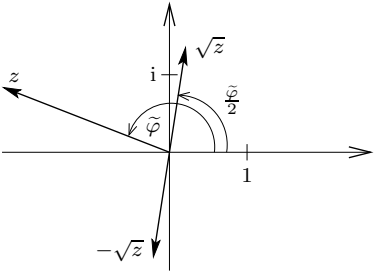}
	\end{picture}%
	\begin{footnotesize}%
	\setlength{\unitlength}{1mm}%
	\begin{picture}(48,36)
		\put(20.5,26.5){\makebox(0,0){i}}
		\put(33,13){\makebox(0,0){1}}
		\put( 2,26){\makebox(0,0){$z$}}
		\put(28,30){\makebox(0,0){$\sqrt{z}$}}
		\put(16, 3){\makebox(0,0){$-\sqrt{z}$}}
		\put(20.5,19.5){\makebox(0,0){$\widetilde{\varphi}$}}
		\put(31,23){\makebox(0,0){$\widetilde{\varphi} \over 2$}}
	\end{picture}%
	\end{footnotesize}%
	\caption{%
	 The square roots $\pm\sqrt{z}$ of a
	 complex number $z$ in the complex plane. If $z$ $=$
	 $\widetilde{r}$ e$^{{\mathrm{i}}\widetilde{\varphi}}$,
	 then $\sqrt{z}$ $=$
	 $\sqrt{\widetilde{r}}$ e$^{{\mathrm{i}}\widetilde{\varphi}/2}.$
	\label{fig-sqrt}}
\end{figure}

\begin{lem}
\label{lemma-tunneling}
	Let $u_s$ be a solution of the differential equation
	$A_{(s)}u_s$ $=$ 0. Then the asymptotic behavior of $u_s$ is
	given by
	$$
		u_s
		\begin{array}{c} \\[-4.25ex]
			\phantom{\scriptstyle r \to \infty}\\[-1.5ex]
			\longrightarrow\\[-1.5ex]
			{\scriptstyle r \to \infty} \\[-1.5ex]
		\end{array}
		a_{\mathrm{in}}\, r^s\, {\mathrm{e}}^{-{\mathrm{i}} \omega r}
		+ a_{\mathrm{out}}\, r^{-s}\, {\mathrm{e}}^{{\mathrm{i}} \omega r},
	$$
	\begin{equation}
		u_s
		\begin{array}{c} \\[-4.25ex]
			\phantom{\scriptstyle r \to M}\\[-1.5ex]
			\longrightarrow\\[-1.5ex]
			{\scriptstyle r \to M} \\[-1.5ex]
		\end{array}
		b_{\mathrm{in}}\, {\mathrm{e}}^{-{\mathrm{i}} k r}
		+ b_{\mathrm{out}}\, {\mathrm{e}}^{{\mathrm{i}} k r}
	\label{u-asymptotic}
	\end{equation}
	with the complex amplitudes
	$a_{\mathrm{in}}$, $a_{\mathrm{out}}$,
	$b_{\mathrm{in}}$, $b_{\mathrm{out}}$ $\in$ ${\mathbb{C}}$ and the
	complex exponent $k$ given by (cf.\ figure \ref{fig-sqrt})
	\begin{equation}
		k^2 = {{ [(M^2+a^2)\omega - am]^2 \over (M^2-a^2)^2 }
		 + { |s| - 1 - \lambda - 4 {\mathrm{i}} s \omega M \over M^2 - a^2 }}\!.
	\label{k-def}
	\end{equation}
\end{lem}
\emph{Proof.}
First we consider the behavior for $r$ $\to$ $\infty$. In this case,
$q_s$ $=$ ${q_s}^{(\infty)}$ $+$ $O(r^{-2})$ with
${q_s}^{(\infty)}$ $=$ $\omega^2$ $-$ $2is\omega/r$. Thus asymptotically
$A_{(s)}u_s$ $\to$ $u_s''$ $+$ ${q_s}^{(\infty)}u_s$, which is solved
by the first limit (\ref{u-asymptotic}) as is seen by direct computing.

On the other hand, we calculate by $K(M)$ $=$
$(M^2+a^2)\omega$ $-$ $am$ that
$q_s$ $\to$ $k^2$
as $r$ $\to$ $M$. Hence,
$A_{(s)}u_s$ $\to$ $u''$ $+$ $k^2 u$, which is solved by
the second limit in (\ref{u-asymptotic}).
\hfill{$\square$}

\phantom{pause}

Note that the complex conjugate of $A_{(s)}$ in (\ref{A-s-def})
equals the radial operator $A_{(-s)}$
of the waves of spin $-s$, $A_{(s)}^*$ $=$ $A_{(-s)}$. In particular,
$A_{(s)}^* u_{-s}$ $=$ $A_{(-s)} u_{-s}$ $=$ 0,
and thus
\begin{equation}
	A_{(s)}^* u_{-s} = A_{(s)} u_{s} = 0.
\label{A*=A=0}
\end{equation}
The advantage of the transformation in theorem \ref{satz-potential} 
is the fact
that the Wronskian of two solutions of the differential equation
$A_{(s)}u_s$ $=$ 0 is constant. In particular, the behavior of two 
\emph{independent}
solutions at the boundaries of the domain of definition determines 
the reflexion and
transmission coefficients of the potential $V_s$, 
cf.\ \cite{Reed-Simon-1979}.
By the peeling theorem of Newman and Penrose \cite{Newman-Penrose-1962}, for a null field of
spin weight $s$ the ingoing and outgoing solutions differ in magnitude by the factor $r^{2s}$.
Accordingly, the solutions $R_s$ show the asymptotic behavior
\begin{equation}
	R_s \propto \underbrace{{1 \over r} \ {\mathrm{e}}^{-{\mathrm{i}} \omega
	  r}}_{\mbox{ingoing waves}}
	\quad \mbox{and} \quad
	  \underbrace{{1 \over r^{2s+1}} \ {\mathrm{e}}^{{\mathrm{i}} \omega r}.}_{\mbox{outgoing waves}}
\label{asymptotic-behavior}
\end{equation}
This is in accordance with the asymptotic behavior of $u_s$,
using the transformation (\ref{u-s-def})
and $\Delta$ $=$ $r^2 + O(r)$.

\begin{satz}\label{satz-tunneling}
	The only massless particles that can tunnel out of a non-extremal
	Kerr black hole are photons with
	the discrete frequencies
	\begin{equation}
		\omega_m = {am \over r_+^2 + a^2}.
	\label{omega-m}
	\end{equation}
	Their wave numbers $k_m$ at $r$ $=$ $M$ are given by
	\begin{equation}
		k_m^2 = {a^2m^2\,(M^2-r_+^2)^2 \over (M^2-a^2)^2 (r_+^2+a^2)^2}
		  - { \lambda + 4 {\mathrm{i}} s \omega_m M \over M^2 - a^2 },
	\end{equation}
	depending on the angular quantum number $m$ $\in$ $\mathbb{Z}$.
	The amplitudes of a photon wave with the asymptotic behavior
	(\ref{u-asymptotic}) is
	determined by the conservation law
	\begin{equation}
		|a_{\mathrm{out}}|^2 - |a_{\mathrm{in}}|^2
		= { {\mathrm{Re}}\, k_m \over \omega_m} \,
		(|b_{\mathrm{out}}|^2 - |b_{\mathrm{in}}|^2),
	\label{conservation-law}
	\end{equation}
\end{satz}
\emph{Proof.}
A Kerr black hole with mass $M$ and rotation parameter $a$ is non-extremal if and only if
$a^2 < M^2$.
For $\omega$ $=$ $\omega_m$ as in (\ref{omega-m}) we compute straightforwardly
the functions $K_m$ $=$ $K$ as
	\begin{equation}
		K_m = \omega_m (r-r_+)(r + r_+).
	\label{K-m}
	\end{equation}
Inserting $\omega_m$ into the expression for $k$ in (\ref{k-def}) yields
$$
		k_m^2 = \left[ {am\,(M^2-r_+^2) \over (M^2-a^2) (r_+^2+a^2)}
			\right]^2
		  + { |s|-1 - \lambda - 4 {\mathrm{i}} s \omega_m M \over M^2 - a^2 }.
$$
By theorem \ref{satz-potential} a massless particle of spin-weight $s$ is given as a wave function
$u_s$ satisfying $A_{(s)}u_s$ $=$ 0.
Since the WKB approximation is applicable if $q_s$ is varying slowly enough,
\begin{equation}
	\left| {1 \over \sqrt{q_s}^3} 
	  {{\mathrm{d}}^2 \sqrt{q_s} \over {\mathrm{d}}r^2} \right| \ll 1
	\qquad \mbox{and} \qquad
	\left| {1 \over q_s} 
	  {{\mathrm{d}} \sqrt{q_s} \over {\mathrm{d}}r} \right| \ll 1
\label{WKB-condition}
\end{equation}
we may use it for a solution $u$ in a region close enough to the
event horizon at $r$ $=$ $r_+$ and far enough from the turning points
$q_s$ $=$ 0, see \cite{Goswami-1997} \S8.2, say
$r_+-\varepsilon_-$ $<$ $r$ $<$ $r_++\varepsilon_+$.
In this regime, the transmission coefficient $|T|^2$ 
is given by
\begin{equation}
	|T|^2 = \exp \left[ -2\int_{r_+-\varepsilon_-}^{r_++\varepsilon_+}
	  {\mathrm{Re}}\,\sqrt{q_s}\ {\mathrm{d}}r \right]
\label{transmission-coefficient}
\end{equation}
The integral diverges if $q_s$ has a pole at $r$ $=$ $r_+$ of order
not smaller than 2, but it attains a finite value for a pole of order 1 since
$$\left|\int_{-\varepsilon_-}^{\varepsilon_+} 
   { {\mathrm{d}}r \over \sqrt{r} }\right|
  \leqq \int_{-\varepsilon_-}^{\varepsilon_+} { {\mathrm{d}}r \over \sqrt{|r|} }
  = 2 (\sqrt{\varepsilon_+} + \sqrt{\varepsilon_-}).$$
There are three terms in $q_s$ which may result in a pole of order 2 at
$r$ $=$ $r_+$, namely 
\begin{eqnarray}
	 \Delta^2 q_s = K^2 + (1-s^2) (r-M)^2 \nonumber \\*
	  {}+ 2 {\mathrm{i}}sK (r-M) + O(\Delta).
	\label{q-s-poles}
\end{eqnarray}
The third term, the imaginary part, vanishes (in the
non-extremal case, where $r_+$ $>$ $M$) if and only if 
$K(r_+)$ $=$ 0 or $s$ $=$ $0$.
The first two terms, the real part, vanish 
at $r$ $=$ $r_+$ if and only if $K^2(r_+)$ $=$ $(s^2$ $-$ $1)$ $(r-M)^2$.
Since $q_s$ is never purely real and negative, a pole both in the real part
as well as in the imaginary part of $q_s$ is also a pole of 
Re\,$\sqrt{q_s}$, cf.\ figure \ref{fig-sqrt}. Thus in (\ref{q-s-poles}) 
the real and the imaginary 
terms of $q_s$ resulting in a pole of order 2 have to vanish simultaneously
at $r$ $=$ $r_+$,
i.e.\ $K(r_+)$ $=$ 0 and $|s|$ $=$ 1.
This means that the integral 
$\int_{M}^\infty {\mathrm{Re}}\,\sqrt{q_s}\, {\mathrm{d}}r$
exists only for photons
with frequencies $\omega=\omega_m$. 
In particular, with (\ref{K-m}) we obtain $K_m/\Delta$ $=$ 
${\omega_m (r+r_+) \over (r-r_-)}$. This leads to
the functions $q_{sm}$ 
$$
	q_{sm} = {\omega_m^2(r+r_+)^2 \over (r-r_-)^2}
	 - { \lambda \over \Delta}
	 + { 2{\mathrm{i}}s \omega_m ( r^2 + r_- r + Mr_+) \over (r-r_+) (r-r_-)^2 }.
$$
Now let $u_s$ be a solution of $A_{(s)}u_s$ $=$ 0 with the asymptotic
behavior (\ref{u-asymptotic}).
Then $u_{-s}$ solves $A_{(-s)}u_{-s}$ $=$ 0 and is therefore a wave of spin $-s$, too.
By $A_{(-s)}^*$ $=$ $A_{(s)}$ it follows that 
$A_{(s)} u_{-s}^*$ $=$ 0,
i.e.\ 
$u_{-s}^*$ 
is a wave with spin $+s$. Hence the Wronskian of the two
solutions of $A_{(s)}u$ $=$ 0 is given asymptotically by
\begin{eqnarray}
	[u_s, u_{-s}^*]
	&
	\begin{array}{c} \\[-4.25ex]
		\phantom{\scriptstyle r \to \infty}\\[-1.5ex]
		\longrightarrow\\[-1.5ex]
		{\scriptstyle r \to \infty} \\[-1.5ex]
	\end{array}
	&
	2 {\mathrm{i}}\omega \, 
	(|a_{\mathrm{out}}|^2 - |a_{\mathrm{in}}|^2),
	\nonumber \\*
	\left[u_s, u_{-s}^*\right]
	&
	\begin{array}{c} \\[-4.25ex]
		\phantom{\scriptstyle r \to M}\\[-1.5ex]
		\longrightarrow\\[-1.5ex]
		{\scriptstyle r \to M} \\[-1.5ex]
	\end{array}
	&
	2 {\mathrm{i}} \ {\mathrm{Re}}\, k\
	(|b_{\mathrm{out}}|^2 - |b_{\mathrm{in}}|^2),
\label{[u,u]-asymptotic}
\end{eqnarray}
Since the Wronskian is constant for an equation of the form $u'' + qu$ 
$=$ 0, the assertion follows.
\hfill{$\square$}

\section{Extremal rotation\label{sec-rotation}}
An extremal Kerr black hole is given by the maximally possible angular momentum $|a|$ $=$ $M$.
In this  case, we have $r_+$ $=$ $r_-$ $=$ $M$, and
the functions $\Delta$ and $K$ in (\ref{rho-Delta}, \ref{K-def}) simplify to
$K$ $=$ $(r^2+M^2) \omega$ $-$ $am$, $\Delta$ $=$ $(r-M)^2$. According to
(\ref{q-s-def}) this yields 
\begin{eqnarray}
	q_s & = &
	 { K^2 \over (r-M)^4 } + { |s| (1- |s|) - \lambda \over (r-M)^2 }
	\nonumber \\* & &
	 {}+ {2 {\mathrm{i}} s \over (r-M)^2 } 
	 \left( 2 \omega r - {K \over (r-M)} \right).
\label{q-s-extremal}
\end{eqnarray}
Now $q_s$ has a pole of order 4 at $r=M$. To apply a similar consideration
as in the proof of theorem \ref{satz-tunneling}, we have to arrange the
nominators to obtain a zero of order 3. For convenience, we again consider the
imaginary and the real part of $q_s$ separately. First we note that $K(M)$ must 
vanish to avoid a pole of order 3, i.e.\ $K(M)$ $=$ 0, or equivalently
\begin{equation}
	\omega_m = { m \over 2a }, \qquad (|a| = M).
	\label{omega-m-extremal}
\end{equation}
With this, $K$ simplifies to $K$ $=$ $(r-M)(r+M)\omega_m$, and therefore
\begin{equation}
	{\mathrm{Im}}\, q_{sm} = { 2s \omega_m \over r-M },
	\label{Im-q-sm-extremal}
\end{equation}
i.e., the imaginary part of $q_s$ has a pole of order 1.
Regarding the real part, we insert our derived expression for
$K$ to obtain
\begin{equation}
	{\mathrm{Re}}\, q_{sm} = 
	 { (r+M)^2 \omega_m^2 + |s| (1- |s|) - \lambda \over (r-M)^2 },
	\label{Re-q-sm-extremal}
\end{equation}
At $r$ $=$ $M$, the nominator vanishes if and only if 
$\lambda$ $=$ $2\omega_m^2 M$ $+$ $|s|(1-|s|)$, or with (\ref{omega-m-extremal}),
\begin{equation}
	\lambda = { m^2 \over 2 } + |s|(1-|s|).
	\label{lambda-sm-extremal}
\end{equation}
Therefore, massless spin waves only can tunnel out of an extremal Kerr black hole
under a restrictive condition on the separation constant $\lambda$. 
Since $|s|(1-|s|)$, 
is integer for integer spin $s$,
$m$ has to be even to guarantee that $m^2/2$ and thus $\lambda$ is an integer.
(Remember that $\lambda$ has to be integer for integer spin $s$.)
On the other hand, $m$ also has to be even such that $\lambda$ is a square
of an odd-half integer for $|s|$ $=$ $\frac12$.
(Note: $|s|(1-|s|)$ $=$ $\frac14$, and
$\lambda$ $=$ $(2m^2 + 1)/4$; since $m^2$ $=$ 0 or 1 mod 4, 
$2m^2+1$ $=$ 1 or 3 mod 4; this is only a square number if $2m^2+1$ $=$ 1 mod 4,
because any square number $n$ $=$ 0 or 1 mod 4.)

\section{Reissner-Nordstr{\o}m black holes\label{sec-RN}}
For $a$ $=$ 0 and 0 $<$ $Q^2$ $\leqq$ $M^2$, the Kerr-Newman spacetime 
represents a static charged black hole, the Reissner-Nordstr{\o}m black hole.
A general master equation like Teukolsky's equation 
governing massless spin waves in the Reissner-Nordstr{\o}m geometry
and leading to radial and angular equations like 
(\ref{radial equation}), (\ref{angular equation})
could not be derived to date.
Nonetheless, the basic equations governing electromagnetic and gravitational
waves ($|s|$ $=$ 1, 2) can be transformed to $T_{(s)}Y_{\pm i}$ $=$ 0, where
$T_{(s)}$ $=$
${{\mathrm{d}}^2\over {\mathrm{d}}r^2}$ $+$ $g_s {{\mathrm{d}}\over {\mathrm{d}}r}$ 
$+$ $f_s$ with
\begin{equation}
		f_s = { r^4 \omega^2 
		 \over \Delta^2 }
		  + { 2 {\mathrm{i}} \omega r \over \Delta }
		  - { n^2 \over \Delta} \left( 1 + { 2 \widetilde{q}_i \over n^2 r } 
			\right)
		    \left( 1 + { 2 \widetilde{q}_j \over n^2 r } \right),
	\label{f-s_RN}
\end{equation}
and
\begin{equation}
		g_s = {2 (r - M) \over \Delta} 
		 + {{\mathrm{d}}\over {\mathrm{d}}r} 
		 \ln \left[ \left({r^6 \over \Delta} \right)^{2|s|-3}
		 \left(1+{2\widetilde{q}_i\over n^2 r}\right) \right].
	\label{g-s-RN}
\end{equation}
Here the separation constant $n$ is given by $n^2$ $=$ $(l-1)(l+2)$, the indices
$i$ and $j$ take the two values $i,j$ $=$ 1,2, but $i\not= j$, and the constants
$\widetilde{q}_i$ are given by
\begin{equation}
	\widetilde{q}_{1/2} = 3M \pm \sqrt{9M^2 + 4 Q n^2},
	\label{q-i-RN}
\end{equation}
These relations can be readily deduced from equations (244) and (245) in
\cite{Chandrasekhar-1983}, \S5 p.~244.

By the same procedure as in the proof of theorem \ref{satz-potential}
the spin-weighted wave equations can be transformed to 
$A_{(s)}P_s$ $=$ 0 with $A_{(s)}$ $=$ d$^2/$d$r^2$ $+$ $q_s$ and
$q_s$ $=$ $f_s$ $-$ $g_s^2/4$ $-$ $g_s'/2$. But since
$\Delta^2 q_s$ $=$ $r^4\omega^2$ $+$ $O(\Delta)$, $q_s$ has a pole
of order not smaller than 2 as long as $\omega>0$. Hence
for waves of spin-weight $|s|$ $=$ 1 or 2 no tunneling
out of a Reissner-Nordstr{\o}m black hole can occur.

\section{The Dirac equation in a Kerr-Newman spacetime\label{sec-Dirac}}
For a particle of mass $m_e$ and electrical charge $e$
let be given the constant
$\mu_e$ $=$ $m_e c/(\sqrt{2}\hslash)$,
such that $\sqrt{2}\pi/\mu_e$ is the Compton wavelength.
Then it is described
by the Dirac equation in a general spacetime, 
cf.\ \cite{de-Vries-1995} eq.~(16)
(for $e=0$).
In the Kerr-Newman geometry it can be reduced
to the complex radial differential equation
$T_{(s)}P_s$ $=$ 0, 
where $T_{(s)}$ $=$ 
${{\mathrm{d}}^2\over {\mathrm{d}}r^2}$ $+$ $g_s {{\mathrm{d}}\over {\mathrm{d}}r}$ 
$+$ $f_s$ with
\begin{eqnarray}
	f_s = { K^2 - 2{\mathrm{i}} s(r-M) K \over \Delta^2 }
	  - { \mu_e K \over 
	  (\lambda + 2{\mathrm{i}}s\mu_e r)\Delta} 
	\nonumber \\*
	  {}+ { 2 {\mathrm{i}} s ( 2 \omega r + eQ) 
	  - \mu_e^2 r^2 - \lambda^2 \over \Delta}
	\label{f-s_KN}
\end{eqnarray}
and
\begin{equation}
	g_s 
	= {r - M \over \Delta}
	  - {2{\mathrm{i}}s\mu_e \over 
		\lambda + 2 {\mathrm{i}}s\mu_e r}
	\label{g-s-KN}
\end{equation}
(note that $P_s$ is related to the usual radial function $R_s$ by
(\ref{radial transformation})), and the angular equation
${\mathcal{L}}S_s = 0$, with 
\begin{eqnarray}
	\lefteqn{%
	{\mathcal{L}} = 
	  {1 \over \sin\theta } {{\mathrm{d}} \over {\mathrm{d}} \theta } 
	  \left( \sin \theta\ {{\mathrm{d}} \over {\mathrm{d}} \theta } \right)
	  + {a \mu_e \sin \theta \over 
		\lambda - 2as \mu_e \cos\theta } 
		{{\mathrm{d}} \over {\mathrm{d}} \theta } 
	}
	\nonumber \\* & &
	  + (a\omega \cos \theta - s)^2
	  + {a \mu_e ( a \omega \sin^2\theta - m - s \cos\theta ) 
		\over \lambda - 2as \mu_e \cos\theta } 
	\nonumber \\* & &
	  - { (m + s \cos\theta)^2 \over \sin^2\theta }
	  - a^2 \mu_e^2 \cos^2 \theta 
	  - s(s-1) + A.
\label{angular Dirac}
\end{eqnarray}
Here
\begin{equation}
	K = K(r) = (r^2 + a^2)\, \omega - am + eQr,
\label{K-def-KN}
\end{equation}
and $\lambda^2 = A + a^2 \omega^2 - 2am\omega$.
The constants
$\lambda$, $A$ $\in$ ${\mathbb{R}}$ are separation constants obtaining
discrete values depending on the boundary conditions of $S_s$.
For details see \cite{Lee-1977},
for the Kerr case ($Q=0$) see \cite{Chandrasekhar-1983}, \S104 
eqs.~(123), (124).

To analyze the potential, a transformation analogous to the proof of 
theorem \ref{satz-potential} has to be done, where now by $4s^2=1$,
\begin{equation}
	{{\mathrm{d}} g_s \over {\mathrm{d}}r} 
	 = { 1 \over \Delta } \left( 1 - {2(r-M)^2 \over \Delta}\right)
	 - { \mu_e^2 \over (\lambda + 2{\mathrm{i}}s\mu_e r )^2 },
\label{g'-s-KN}
\end{equation}
cf.\ (\ref{g'-s}).
This yields a complex potential $q_s$
differing from (\ref{q-s-def}) only by terms depending on $\mu_e$. 
As long as 
$\lambda$ $\not=$ $-2{\mathrm{i}}s\mu_e r_+$, these
terms are of order $O(\Delta^{-1})$, and
the proof of theorem \ref{satz-potential} in essence can be applied.
Hence no tunneling of electrons out of a Kerr-Newman black 
hole can occur, unless for the case of extremal rotation. 

If $\lambda$ $=$
$-2{\mathrm{i}}s\mu_e r_+$
the situation changes, since now the terms by which $q_s$ differs from
the massless case, are  of order $O(\Delta^{-2})$. 
First we notice that
$\lambda$ $+$ $2{\mathrm{i}}s\mu_e r$ $=$ 
$2{\mathrm{i}}s\mu_e (r-r_+)$. Therefore, with 
$\Delta$ $=$ $(r-r_+)(r-r_-)$,
$4s^2$ $=$ 1, 
$4s^2(M-r) + r - r_-$ $=$ $M-r_-$,
we obtain
\begin{eqnarray}
		f_s & = & { K^2 \over \Delta^2 }
		  + { {\mathrm{i}} K ( M - r_- ) \over 2s\Delta^2 }
		  + { {\mathrm{i}} ( 2 \omega r + eQ)
			\over 2s\Delta}
		  - { r + r_+ \over r - r_- } \, \mu_e^2,
	\nonumber \\*
	g_s & = & {}- {M - r_- \over \Delta},
	\label{f-g-s-KN-special}
\end{eqnarray}
With these equations the terms deciding the occurence of a pole
of order 2 at the horizon $r$ $=$ $r_+$ are determined by
\begin{eqnarray}
	\Delta^2 q_s = K^2 + (M - r_-) \left(r + {r_- - 5M \over 4} \right) 
	\nonumber \\*
	  + 2{\mathrm{i}}s K (M-r_-) + O(\Delta).
\end{eqnarray}
To avoid a pole of order 2, the imaginary part has to vanish. This 
implies $K(r_+)=0$ (even in the extremal case $M$ $=$ $a^2$ $+$ $Q^2$,
for then $\Delta$ $=$ $(r-M)^2$ and $M$ $=$ $r_-$), or
\begin{equation}
	\omega_{m} = { am - eQr_+ \over r_+^2 + a^2 }.
	\label{K=0-KN}
\end{equation}
The real part
$\Delta^2$ $\cdot$ Re\,$q_s$ $=$ $(M - r_-) (r - \frac14 (r_- - 5M))$
then vanishes if and only if $r_+$ $=$ $r_-$ $=$ $M$, 
which is equivalent to
$M^2$ $=$ $a^2$ $+$ $Q^2$. This is the extremal case, and we see that
then $K$ $=$ $\omega_{m}$ $(r-M)(r+M)$, i.e.\ $q_s = f_s$ or
\begin{equation}
	q_s = \omega_{m}^2 \ {(r+M)^2 \over (r-M)^2} 
		  + { {\mathrm{i}} ( 2 \omega_{m} r + eQ)
			\over 2s (r-M)^2}
		  - { r + M \over r - M } \, \mu_e^2.
\end{equation}
Therefore, $q_s$ has a pole of order 2 at the event horizon, such that no tunneling
may occur for $\lambda$ $=$ $-2{\mathrm{i}}s\mu_e r_+$.

\section{Discussion\label{sec-discussion}}
In the present paper we 
consider 
properties of spin-weighted waves in various Kerr-Newman spacetimes.
We analyze the singular 
potential barriers of the respective wave equations emerging at the event horizon. It turns out that the potential barriers are repulsive enough
to prevent spin waves with spin $|s|$ $=$ 0, $\frac12$, 2 
from tunneling out of the black hole, but weak enough to admit
the tunneling of photons in case of non-extremal rotation.

In particular, if $a$ denotes the angular momentum of a Kerr-Newman
spacetime and $Q$ its electrical charge, then the following table
shows the occurence of tunneling (`$\times$'), its impossibility (`--') and
the present ignorance (`?') with respect to the various kinds of
spins $s$.
\begin{equation}
\mbox{%
\begin{tabular}{l@{\quad}cccc}
	\hline \hline \\[-2ex]
	spin $|s|$ & $0$ & $\frac12$ &   $1$   &   $2$ \\[.5ex] \hline
	$a=0$ &   --     &    --     &   --    &   --    \\
	$0<|a|<M$, $Q=0$
	      &   --     &    --     & $\times$&   --    \\
	$|a|=M$, $Q=0$
		 & $\times$& $ \times$  & $\times$& $\times$ \\
	$0<|a|, |Q|<M$
	      &   --     &    --     &    ?    &    ?    \\
	\hline \hline
\end{tabular}%
}
\end{equation}
The case $a\not=0\not=Q$, i.e.\ charged rotating Kerr-Newman black holes,
still remains unknown. The crucial role that rotation plays in the known 
cases suggests that photons may tunnel out of \emph{any} rotating
black hole. A very strong hint supporting this conjecture is the analysis
of one of the authors \cite{de-Vries-1996}, where any electromagnetic
wave \emph{in the outer space} of a rotating Kerr-Newman black hole
can gain some energy from the hole. However, we still have no such hints
for gravitons.

For the case of extremal rotation, $|a|=M$, any particle can tunnel 
out of the black hole. Because of the cosmic censorship hypothesis 
(\ref{ccc}),
$Q$ vanishes necessarily in this case.
There is only the restriction on the separation constant 
$\lambda$ given by (\ref{lambda-sm-extremal}) to be an even integer.
Whether such an integer exists at all depends on the Sturm-Liouville 
equation for $S_s$, (\ref{angular equation}) or (\ref{angular Dirac}),
respectively.

\begin{figure} 
	\centering
	\includegraphics[width=\columnwidth]{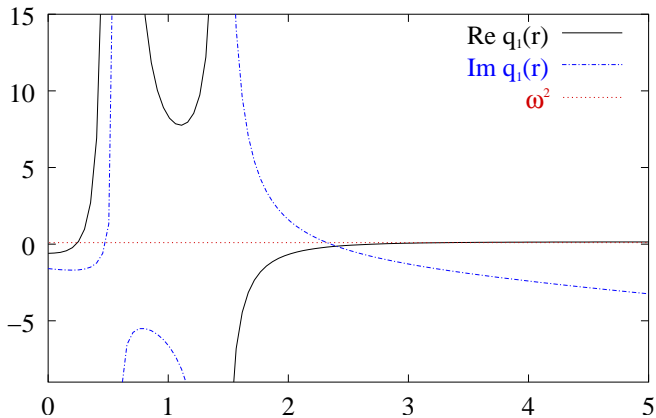}
	\caption{\label{fig-pot} 
	The real and imaginary part of $q_s$ for $a=.9$, $Q=0$,
	$s=1$, $m=1$, $\omega=\omega_m$; 
	also plotted is the constant $\omega_m^2$.
	The tics on the horizontal $r$-axis are in units of $M$.
	Note the poles at $r_+$ $=$ $1.436M$ and $r_-$ $=$ $.564M$.} 
\end{figure}

Mathematically, in case of $Q=0$ or $a=0$ a
necessary and sufficient condition for the
occurence of the tunnel phenomenon is the vanishing of $g_s$,
the factor of the first derivative in the radial wave equations,
and of the function $K$,
cf.\ equations (\ref{g-s}), (\ref{K-def}) for massless waves of spin $s$, 
and (\ref{g-s-KN}), (\ref{K-def-KN}) for spin-$\frac12$ waves of mass 
$m_e$ $=$ $\sqrt{2} \hslash \mu_e / c$. This condition yields
a potential with an order-1 pole at the event horizon 
$r$ $=$ $r_+$, see figure \ref{fig-pot}.
Remarkably, the real part of a potential with a pole of order one is 
attractive from the outer space of the black hole
and repulsive from the interior, but the imaginary part tends
to $+\infty$ from the outside and to $-\infty$ from the interior. 
A wave packet inside the black hole with nonvanishing outgoing
amplitude $b_{\mathrm{out}}$ then reaches the outer space with 
nonvanishing amplitude $a_{\mathrm{out}}$; 
if, e.g., $b_{\mathrm{out}}$ is known at $r=M$ and 
$b_{\mathrm{in}}=0$, the outside
amplitude is given by (\ref{conservation-law}) with
$a_{\mathrm{in}}=0$.

Where does this all lead us to?
Notably, the tunnel phenomenon studied in this paper is a 
semiclassical phenomenon. This implies that the black hole gravitational
field has to be strong enough compared to the typical energy of the
wave field such that it does not perturb the curvature essentially,
i.e.\ such that it is a test field.

The tunnel phenomenon seems to be related to another semiclassical
phenomenon, the Hawking radiation. But whereas the Hawking radiation
is a black hole thermodynamical effect occuring in particular for
static black holes, the tunneling only exists in case of rotation.
As such it resembles the classical superradiance effect,
but this affects only, and all, particles with integer spin.

A complete theory of quantum gravity must include the tunnel 
phenomenon --- and should
explain it. Especially the puzzles 
concerning gravitons and photons in charged rotating black holes
should be solved.
To date, the most promising candidate for a theory of quantum 
gravity is M-theory. It has made strong progress in the last few years,
not only with respect to black hole phenomena. One of the most
remarkable aspects in this context certainly is the deduction of
the notion of entropy of a black hole \cite{Strominger-Vafa-1996}.
However, to our knowledge the analysis of rotating black holes, be they
uncharged or charged, in
M-theory is still missing. 

To conclude, rotation in general relativity over and over again
reveals surprising phenomena,
a fact which is not only demonstrated by
\cite{Abramowicz-1990}, 
\cite{de-Vries-Schmidt-Kaler-1999},
or 
\cite{de-Vries-2000}.
Succeeding in integrating rotation in M-theory would give deep insight.


\end{document}